\documentclass[letterpaper,useAMS,usenatbib]{mn2e}

\usepackage{lscape}
\usepackage{graphicx}
\usepackage{natbib}
\usepackage{graphics}
\usepackage{rotating,multirow}
\usepackage{subfigure}
\usepackage{colortbl}
\usepackage{hyperref}
\usepackage{amssymb,multirow}

\def \vhel{\ifmmode{~V_{{\rm HEL}}}\else{~$V_{{\rm HEL}}$}\fi}
\def \vsys{\ifmmode{~V_{{\rm SYS}}}\else{~$V_{{\rm SYS}}$}\fi}
\def \HA {\ifmmode{{\rm\H}\alpha}\else{${\rm\ H}\alpha$}\fi}

\def \msun{\ifmmode{{\rm\ M}_\odot}\else{${\rm\ M}_\odot$}\fi}

\def \myr{\ifmmode{{\rm\ M}_\odot{\rm\ yr}^{-1}}
        \else{${\rm\ M}_\odot$ yr$^{-1}$}\fi}

\def \mdot{\ifmmode{\dot{M}}\else{$\dot{M}$}\fi}
\def \tena#1 #2 {\ifmmode{#1 \times 10^{#2}}\else{$#1 \times 10^{#2}$}\fi}
\def \kms{\ifmmode{~{\rm km\,s}^{-1}}\else{~km s$^{-1}$}\fi}

\def \apj{ApJ}
\def \mnras{MNRAS}

\def \araa{ARA\&A}
\def \aap{A\&A}

\def \apjl{ApJL}
\def \nat{Nat}

\def \atel{Astron. Telegram}

\title[A weak compact jet in a soft state of Cygnus X-1]{A weak compact jet in a soft state of Cygnus X-1}
\author[A. Rushton et al.]
{A. Rushton,$^{1,2}$\thanks{E-mail: anthony.rushton at eso.org (AR)}
J.C.A. Miller-Jones,$^{3}$ R. Campana,$^4$ Y. Evangelista,$^4$ Z. Paragi,$^{5,6}$ \newauthor T.J. Maccarone,$^7$ G.G. Pooley,$^8$
V. Tudose,$^{9,10,11}$ R.P. Fender,$^7$ R.E. Spencer,$^{12}$  \newauthor and V. Dhawan$^{13}$\\
$^1$European Southern Observatory, Karl-Schwarzschild-Str 2, 85748 Garching, Germany\\
$^2$Onsala Space Observatory, SE-439 92, Sweden\\
$^3$ICRAR - Curtin University of Technology, GPO Box U1987, Perth, WA 6845, Australia\\
$^4$INAF/IASF-Roma, Via Fosso del Cavaliere 100, Roma, I-00133 Italy\\
$^5$Joint Institute for VLBI in Europe, Postbus 2, 7990 AA Dwingeloo, The Netherlands\\
$^6$MTA Research Group for Physical Geodesy and Geodynamics, P.O. Box 91, H-1521 Budapest, Hungary\\
$^7$School of Physics and Astronomy, University of Southampton, Highfield, Southampton SO17 1BJ\\
$^8$Cavendish Laboratory, J. J. Thomson Avenue, Cambridge CB3 0HE\\
$^{9}$ASTRON, Oude Hoogeveensedijk 4, 7991 PD Dwingeloo, the Netherlands\\
$^{10}$Astronomical Institute of the Romanian Academy, Cutitul de Argint 5, RO-040557 Bucharest, Romania\\
$^{11}$Research Center for Atomic Physics and Astrophysics, Atomistilor 405, RO-077125 Bucharest, Romania \\
$^{12}$Jodrell Bank Centre for Astrophysics, School of Physics and Astronomy, University of Manchester, M13 9PL\\
$^{13}$NRAO Domenici Science Operations Center, 1003 Lopezville Road, Socorro, NM 87801, USA\\
}

\begin{document}

\date{Accepted 2011 October 05. Received 2011 October 05; in original form 2011 July 22}

\pagerange{\pageref{firstpage}--\pageref{lastpage}} \pubyear{YYYY}

\maketitle

\label{firstpage}

\begin{abstract}
We present evidence for the presence of a weak compact jet during a soft X-ray state of Cygnus X-1. Very-high-resolution radio observations were taken with the VLBA, EVN and MERLIN during a hard-to-soft spectral state change, showing the hard state jet to be suppressed by a factor of about $3-5$ in radio flux and unresolved to direct imaging observations (i.e. $\lesssim 1$~mas at 4~cm). High time-resolution X-ray observations with the RXTE-PCA were also taken during the radio monitoring period, showing the source to make the transition from the hard state to a softer state (via an intermediate state), although the source may never have reached the canonical soft state. Using astrometric VLBI analysis and removing proper motion, parallax and orbital motion signatures, the residual positions show a scatter of $\sim0.2$~mas (at 4~cm) and $\sim3$~mas (at 13~cm) along the position angle of the known jet axis; these residuals suggest there is a weak unresolved outflow, with varying size or opacity, during intermediate and soft X-ray states. Furthermore, no evidence was found for extended knots or shocks forming within the jet during the state transition, suggesting the change in outflow rate may not be sufficiently high to produce superluminal knots.
\end{abstract}


\section{Introduction}

The outflow of highly collimated jets appears to be a universal aspect of accreting black holes (BH), from stellar-mass X-ray binaries (XRBs) to supermassive active galactic nuclei. Galactic XRBs are at the low end of this mass range, and due to their relative proximity and rapid changes in mass accretion rate ($\dot{m}$), their outburst cycles can be well observed over a few weeks. Therefore, understanding the disk-jet relationship in each of the characteristic accretion states of an XRB is of critical importance in developing a universal (scale invariant) relationship holding over all BH masses.

BH XRBs broadly accrete in two states that are classified according to their X-ray spectra, although many intermediate states have also been identified. \textit{Soft X-ray states} are seen only at high X-ray luminosities \linebreak($\ga1\%$~L$_{\rm{edd}}$) dominated by a thermal X-ray component, and are associated with weak radio emission. \textit{Hard X-ray states} can show all luminosities between $1\gtrsim L/L_{\rm edd}\gtrsim10^{-8}$ and are dominated by a power-law component extending to $\sim100$~keV; this state is also associated with relatively strong, flat-spectrum radio emission. Compact jets are normally inferred from the flat spectrum, although only two BH XRBs have shown resolved jets (using VLBI imaging) during hard states: GRS\,1915+105~\citep{2000ApJ...543..373D} and \break Cygnus X-1~\citep{2001MNRAS.327.1273S}.

The exact disk-jet coupling for XRBs is therefore related to the particular accretion state. During hard states, the power-law component is attributed to thermal Comptonisation by a hot plasma in an optically thin corona around the inner accretion disk, which acts as a ``reservoir'' for the jet outflow~\citep{1975ApJ...195L.101T,1976SvAL....2..191B,1979Natur.279..506S}. When the mass accretion rate increases, the disk temperature rises and the bolometric luminosity starts to become dominated by the thermal disk. It has been suggested that the hard-to-soft state transition is also associated with a rapid increase in the jet Lorentz factor, until the position of the source in the X-ray hardness-intensity diagram (HID) crosses the `jet-line'~\citep{2004MNRAS.355.1105F}, when shocks may be formed within the jet, observed as discrete ejecta propagating away from the system~\citep{2003ApJ...597.1023V}. Once in the soft state, the outflowing material is thought to be quenched and no compact core jets have previously been detected -- see observations \citep[by][]{1972ApJ...177L...5T,1999MNRAS.304..865F} and theoretical work \citep[by][]{1999ApJ...512..100L, 2001ApJ...548L...9M}.

\begin{table*}
\footnotesize
\caption{\label{table:journal}Details of the radio observations of Cygnus X-1 taken in 2010 July. Orbital phase is defined as the superior conjunction of the black hole \citep[using the ephemeris in][]{1999A&A...343..861B}.}
\begin{center}\begin{tabular}{ccccccccccc}
\hline
VLBI & Epoch & Orbital & Start & End & Array & $\lambda$  & $I_{\rm{total}}$   & Noise &  $\theta_{\rm{maj}}$$\times$$\theta_{\rm{min}}$ & PA \\
obs. & (MJD) & phase &  (UT) &  (UT)                 &          &     (cm)         &   (mJy)          & ($\mu$Jy/bm)    &   (mas) & degrees \\
\hline 
\hline
 &55\,380.1 & 0.840 & 02--19:38 & 03--09:28 & MERLIN & 6 &  $5.5\pm0.2$ &  $114$ & 59.9$\times$40.5 & 28 \\
 &55\,381.0 & 0.001 & 03--16:30 & 04--09:28 & MERLIN & 6 &  $5.3\pm0.2$ &  $122$ & 50.9$\times$43.7 & 5 \\
 &55\,382.0 & 0.179& 04--16:31 & 05--09:28 & MERLIN & 6 &  $10.4\pm0.2$ &  $124$ & 51.5$\times$43.7 & 7 \\
 &55\,383.1 & 0.376 & 05--17:30 & 06--09:20 & MERLIN & 6 &  $6.5\pm0.2$ &  $113$ & 58$\times$50 & 27 \\
0 & 55\,386.0 & 0.893 & 08--18:32 & 09--05:39 & e-EVN & 6 &  $15.2\pm0.1$ &  $116$ & 13.8$\times$9.6 & -70 \\
 &55\,387.3 & 0.126 & 09--20:39 & 10--09:28 & MERLIN & 4 &  $6.7\pm0.2$ &  $106$ & 88$\times$38 & -25 \\
 &55\,388.0 & 0.251 & 10--16:30 & 11--09:28 & MERLIN & 4 &  $2.3\pm0.3$ &  $148$ & 136$\times$82 & -4 \\
1 & 55\,388.0  & 0.251 & 10--18:17 & 11--05:24 & e-EVN & 6 &  $5.0\pm0.1$ &  $97$& 13.6$\times$7.8 & -73\\
 &55\,388.8 & 0.394 & 11--16:31 & 11--21:23 & MERLIN & 4 &  $2.4\pm0.2$ &  $132$ & 130$\times$80.0 & -20 \\
\multirow{2}{*}{2} & \multirow{2}{*}{55\,389.4} & \multirow{2}{*}{0.501}& \multirow{2}{*}{12--08:45} & \multirow{2}{*}{12--11:15} & \multirow{2}{*}{VLBA} & 4 &  $5.9\pm0.2$ &  $155$ & 2.2$\times$0.8 & -23\\
&&&&& & 13 & $5.3\pm0.2$ & 184 & 9.8$\times$8.1 & 32\\
\multirow{2}{*}{3} & \multirow{2}{*}{55\,392.4} & \multirow{2}{*}{0.037}&\multirow{2}{*}{15--06:28} & \multirow{2}{*}{15--11:33} & \multirow{2}{*}{VLBA} & 4 &  $3.8\pm0.1$ &  $104$ & 2.0$\times$0.7& -16\\
&&&&&& 13 & $5.2\pm0.2$ & 157 & 15.0$\times$8.0 & -34\\
\multirow{2}{*}{4} & \multirow{2}{*}{55\,394.4} & \multirow{2}{*}{0.394} &\multirow{2}{*}{17--06:23} & \multirow{2}{*}{17--11:33} & \multirow{2}{*}{VLBA} & 4 &  $3.1\pm0.1$ &  $102$ & 2.1$\times$0.7 & -18\\
&&&&&& 13  & $3.1\pm0.2$ & 172 & 15.1$\times$7.9 & -35 \\
\multirow{2}{*}{5} & \multirow{2}{*}{55\,396.4} &\multirow{2}{*}{0.751} &\multirow{2}{*}{19--06:27} & \multirow{2}{*}{19--11:34} & \multirow{2}{*}{VLBA} & 4 &  $1.4\pm0.1$ &  $95$ &2.1$\times$0.7 & -20\\
&&&&&& 13  & $1.2\pm0.2$ & 159 & 15.4$\times$8.0 & -36 \\
\multirow{2}{*}{6} & \multirow{2}{*}{55\,399.4} &\multirow{2}{*}{0.287} &\multirow{2}{*}{22--06:25} & \multirow{2}{*}{22--11:33} & \multirow{2}{*}{VLBA} & 4 &  $2.7\pm0.1$ &  $99$ &1.9$\times$0.7 & -16\\
&&&&&& 13  & $5.7\pm0.2$ & 206 & 20.0$\times$7.3 & 45 \\
 &55\,400.1&0.412 & 23--00:30 & 23--05:30 & WSRT & 6 &  $3.2\pm0.2$ &  $50$& 16000$\times$6000 & -36\\
\hline
\end{tabular}\end{center}
\end{table*}

We present a detailed, high-resolution VLBI monitoring campaign of the XRB Cygnus X-1 over a hard-to-soft state transition. This system contains a black hole candidate with a mass of $14.8\pm1.0\msun$ in a 5.6-day orbit with a supergiant star of mass $19.2\pm1.9\msun$ \citep{2011arXiv1106.3689O,1987SvA....31..170K}. Mass transfer between the star and compact object occurs via a focussed stellar wind from the supergiant. During the hard state, \cite{2001MNRAS.327.1273S} showed that at 8.4~GHz, a $\sim15$~mas jet was persistently present to the north-west (N-W) of the core over the full binary orbit, with a position angle of $\sim22^{\circ}$ west of north. More recent VLBI observations in 2009 confirmed that the compact jet was still present with the same orientation~\citep{2011arXiv1101.3322R}. \cite{2006MNRAS.369..603F} presented evidence that a weak transient knot was produced $\sim50$~mas from the core of Cygnus X-1, after the X-ray state crossed a `jet-line' during a hard-to-soft state transition in 2004.  However, their observations were limited by poor {\it uv}-coverage. Our aims were to verify the existence of these knots, determine whether the compact jet is completely quenched in the soft state, and test the existence of the `jet-line'.

\begin{figure*}
\centering 
\includegraphics[width=16cm]{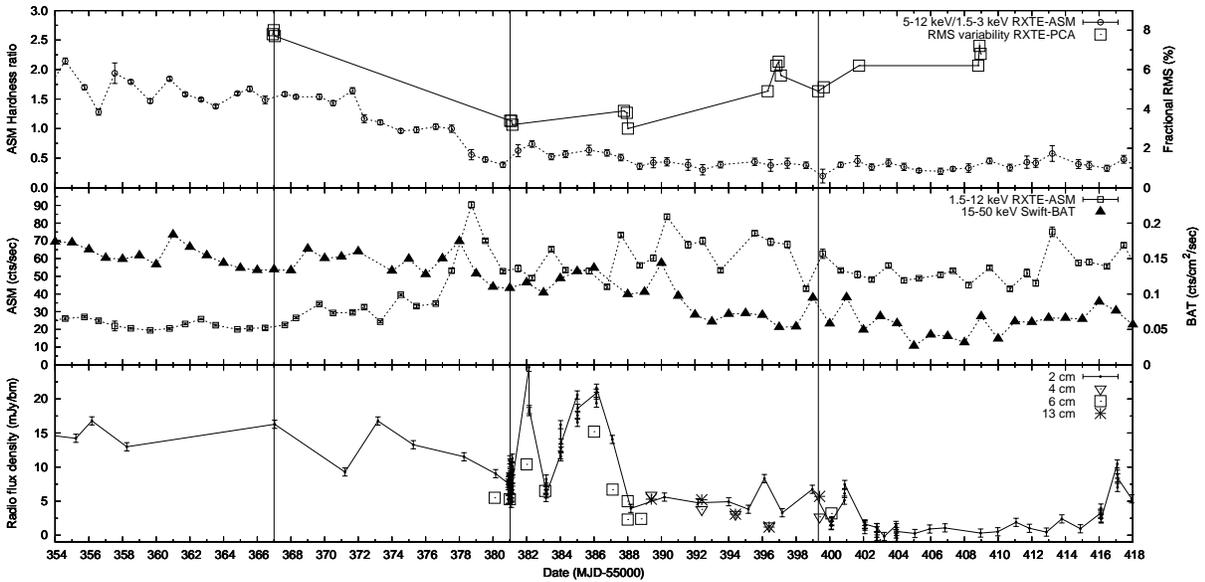} 
\caption{X-ray and radio light curves of Cygnus X-1 over a hard-to-soft state transition during 2010 June/July. From top to bottom: The fractional RMS variability of the X-ray emission monitored with the RXTE-PCA and the $5-12~\rm{keV}/1.5-3~\rm{keV}$ hardness ratio from the RXTE-ASM; total X-ray intensity between $1.5-12$~keV and $15-50$ keV with the RXTE-ASM and Swift-BAT respectively; AMI-LA radio flux density at 2~cm, including the epochs and radio flux densities measured by the e-EVN, MERLIN, VLBA and WSRT (at 4, 6 or 13~cm). The solid vertical lines mark the RXTE-PCA power spectra shown in Fig.~\ref{fig:XTE_results}.} 
\label{fig:lightcurve} 
\end{figure*} 

\section{Observations and results}

Targeted radio and X-ray observations of Cygnus X-1 were triggered after a series of \textit{Astronomer's Telegram}\footnote{\href{http://www.astronomerstelegram.org}{www.astronomerstelegram.org}}(ATel) reports suggested a hard-to-soft state transition was starting in late 2010 June \citep[e.g.][]{2010ATel.2711....1N,2010ATel.2714....1R,2010ATel.2715....1S}. Table~\ref{table:journal} gives a summary of all long baseline radio observations and Fig.~\ref{fig:lightcurve} shows the radio (AMI-LA, MERLIN, e-EVN, VLBA and WSRT) and X-ray (RXTE-ASM and Swift-BAT)  light curves.

\subsection{Radio monitoring}

Initial high-resolution radio monitoring was carried out with MERLIN at 6 and 4~cm (between 2010 July 2 and 10) followed by VLBI observations with the EVN (at 6~cm) on 2010 July 8 and 10 in e-VLBI mode (a.k.a. e-EVN); participating e-EVN telescopes were Jodrell Bank MkII, Knockin, Cambridge, Westerbork, Effelsberg, Torun, Yebes, Medicina, Onsala 25-m and Shanghai. Data were transferred from each antenna to the correlator using high-speed dedicated network links, sustaining connection rates of up to 1024~Mbps per antenna, yielding maximum bandwidths of 128~MHz per polarisation, with dual polarisation. Follow up VLBA observations were scheduled in the dual 4/13 cm mode, using the long wavelength to probe larger spatial scales. A total of five VLBA epochs were taken, on 2010~July 12, 15, 17, 19 and 22 using all 10 antennas, with a recording rate of 512 Mbps (divided equally between 4 and 13~cm), giving a total bandwidth for each frequency band of 32~MHz per polarisation, with dual polarisation. All VLBI observations were phase referenced to the calibrator J1953+3537, separated by a distance of $1.1^{\circ}$ from the target ($1.0^{\circ}$ in RA and $0.4^{\circ}$ in Dec.). In the last four VLBA observations a geodetic VLBI calibrator block was inserted, in order to improve the astrometric accuracy and image quality of the target source without using self-calibration (see \textsc{aips}  Memo 110 for more details). Also we substituted J1957+3338 for every $\sim7^{\rm{th}}$ scan on the target source for a positional check. The VLBA data were correlated using the NRAO implementation of the DiFX software correlator \citep{Del11}. All data were reduced using the standard \textsc{aips} VLBI algorithms (e.g. \textsc{vlbautil}), using the standard EVN pipeline for initial processing of the e-EVN data. Finally a single lower resolution observation was taken with the WSRT on 2010 July 23.

\begin{figure}
\centering 
\includegraphics[width=6.5cm]{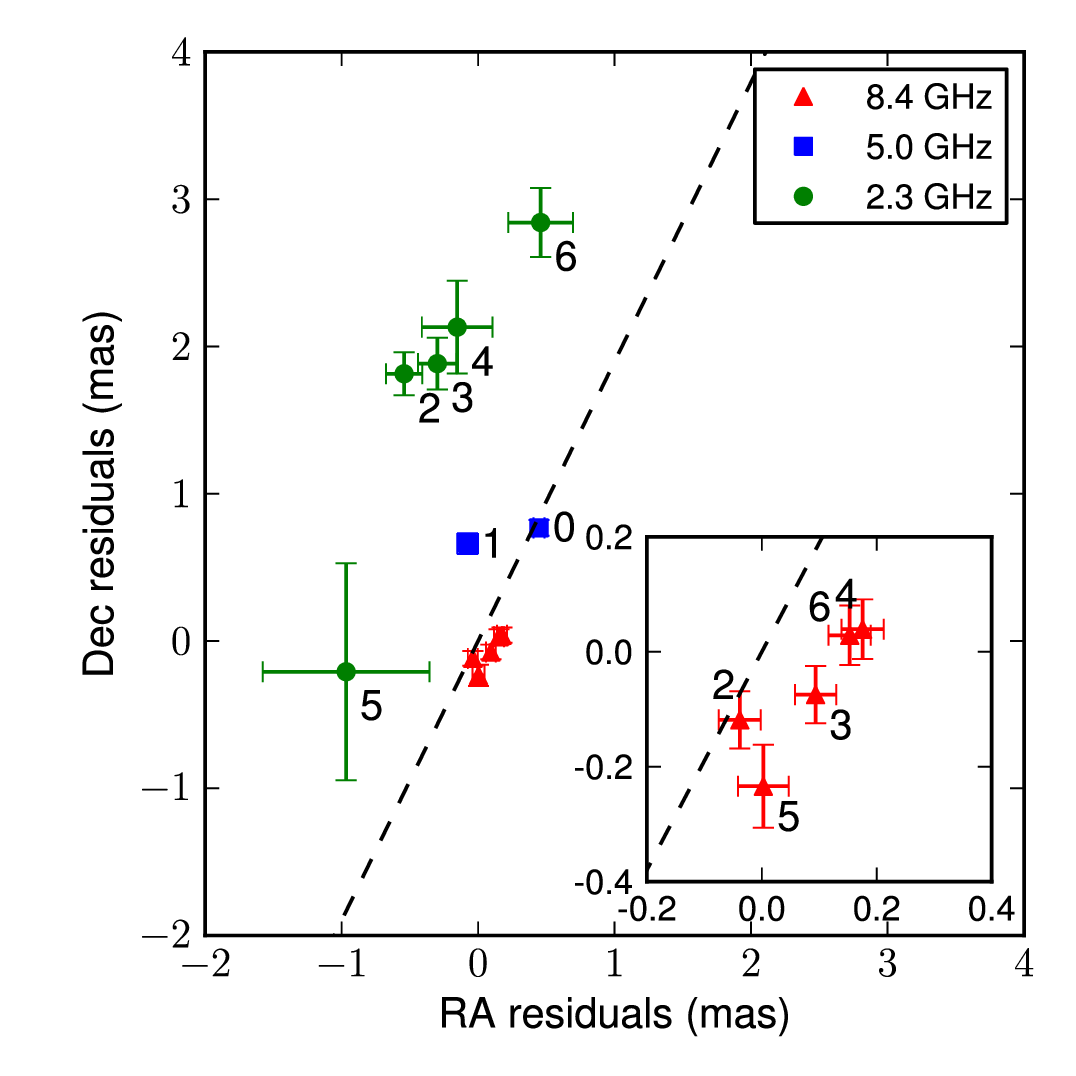} 
\caption{Residual astrometric VLBI positions of Cygnus X-1 in 2010 July, after removing the proper motion, parallax and orbital motion signatures. Note the longer wavelength positions are systematically offset due to a frequency dependence of the fitted position of the phase reference calibrator. Epochs 0--2 had no geodetic calibration block, possibly causing a slight offset.} Inset shows a zoom-in of the 4-cm positions. Numbers indicate the ordering of the 2010 VLBI epochs listed in table~\ref{table:journal}. Dashed black line indicates the known jet axis.
\label{fig:VLBI_results} 
\end{figure}

All high-resolution radio observations of Cygnus X-1 were found to be unresolved down to the beam size of the respective arrays, showing that the compact core jet was quenched to less than a few milliarcseconds in size. Also despite clear evidence of a typical X-ray spectral state change (Section~\ref{sec:X-ray}), no evidence for discrete ejecta was found immediately after the transition. Although the source was unresolved, we fitted the position of the centroid of the emission at each epoch and subtracted the estimated contributions of proper motion, parallax and orbital motion \citep{2011arXiv1106.3688R}. The residual positions are shown in Fig.~\ref{fig:VLBI_results}; estimates of the systematics are $\sim38~\mu$arcsec in RA and $\sim47~\mu$arcsec in Dec. \citep{2006A&A...452.1099P}, which was confirmed by scaling the measured scatter in the check source positions ($39~\mu$arcsec in RA and $55~\mu$arcsec in Dec. at 4~cm) by the relative distances of the target and check source to the phase reference calibrator. Also note that an error was reported in the \textsc{aips} Earth Orientation Parameter (EOP) affecting all VLBI astrometric experiments between 2009 Sept. 21 and 2011 Aug. 04; however, after re-analysing the data with the corrected EOPs we found a positional shift of only $18\pm35~\mu$arcsec at X-band (i.e. within the error).

At both 4 and 13~cm, the residuals are scattered along an axis which is aligned with the known position angle of the hard state jet (note that epochs 0--2 did not have a geodetic calibration block, possibly causing a slight offset). The positional shift between different frequencies is an artefact caused by a frequency and resolution dependent shift in the fitted centroid position of the phase reference calibrator, as a jet structure is seen to the south of the core at 4~cm in J1953+3537 (with a PA of $-172^{\circ}$ east of north) that is unresolved at 13~cm. This shifts the measured centroid at 13~cm along the calibrator jet axis, away from the X-band core, from which the assumed source position was taken. The alignment of our astrometric residuals with the known jet axis in Cygnus X-1 is consistent with the existence of an unresolved compact jet during our observations, with slightly varying size or optical depth ($\tau$).

Cygnus X-1 has been continuously monitored at lower angular resolution by the AMI-LA at 2~cm (also shown in Fig.~\ref{fig:lightcurve}). A clear dip or quenching of the radio emission was observed in correlation with the X-ray spectral state change. During the VLBI observations, AMI-LA also detected radio emission, which suggested the source to have a flat or slightly inverted spectrum. However, at a later stage in the evolution of the soft state (between 2011~February~08 and 2011~April~09) the 2~cm flux became even weaker and the AMI-LA did not detect Cygnus X-1 (the formal mean flux density was $-6 \pm 80~\mu$Jy).

\subsection{X-ray monitoring}
\label{sec:X-ray}

\begin{figure*}
\centering 
\includegraphics[width=12.0cm]{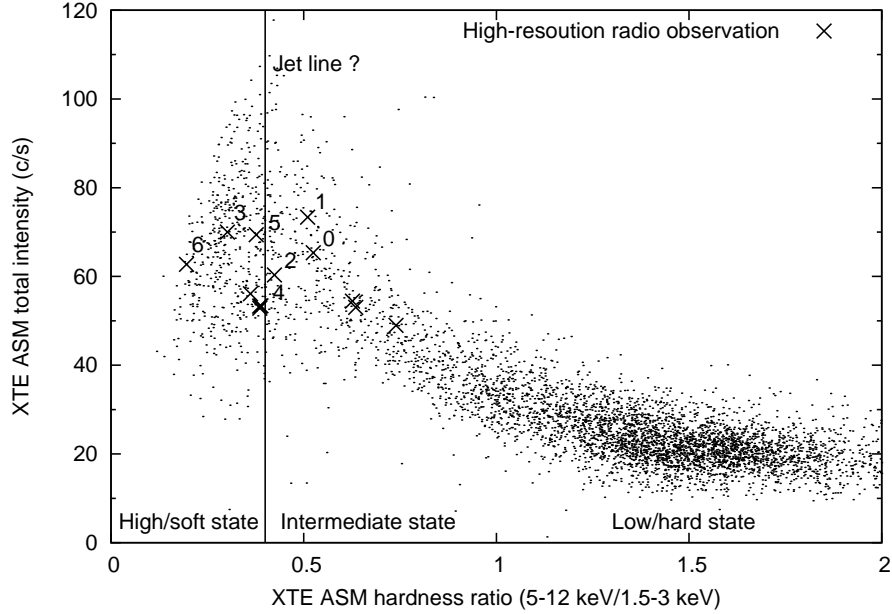} 
\caption{The HID of Cygnus X-1 from RXTE-ASM observations from 1996--2011. The $\frac{5-12~\rm{keV}}{1.5-3~\rm{keV}}$ ``hardness ratio" is compared with the total $1.5-12$~keV X-ray intensity (c/s) binned into one day averages. Crosses mark the closest RXTE-ASM epoch to each observation taken with either MERLIN, e-EVN, VLBA or WSRT (with the VLBI epochs numbered), showing that the proposed `jet-line' was crossed during the high-resolution radio monitoring.} 
\label{fig:HID}
\end{figure*}

In order to analyse the precise X-ray state of Cygnus X-1, targeted fast timing X-ray observations were triggered during the high-resolution radio monitoring, as well as using the RXTE-ASM and Swift-BAT monitoring instruments. A total of 19 pointed RXTE-PCA observations were carried out between 2010 June 19 and 2010 July 31, corresponding to about 68.5~ks of net exposure. All the observations were performed in the binned data mode (\textsc{B\_2ms\_8B\_0\_35\_Q}), with 1.95~ms bin size in the $\sim2.1-14.8$~keV energy band. Data analysis was carried out using custom IDL software. Using the entire energy band we extracted the power spectra normalised to units of fractional squared root-mean-square (RMS) and the fractional RMS variability of the whole dataset. A HID of Cygnus X-1 from the RXTE-ASM over the period 1996--2011 is also shown in Fig.~\ref{fig:HID}. The $1.5-12$~keV X-ray intensity (binned into one day averages) is plotted against a hardness ratio $\left( \rm{HR}=\frac{5-12~\rm{keV}}{1.5-3~\rm{keV}}\right) $. Most of the RXTE-ASM observations over the 15 year period were taken in the hard state, while the VLBI observations presented here were taken during a transition between the intermediate and soft states (as marked by crosses in Fig.~\ref{fig:HID}). It has been proposed by~\cite{2006MNRAS.369..603F} that the `jet-line' is approximately at HR~$\sim0.4$, hence our observations sampled this period well.

\begin{figure}
\centering 
\includegraphics[width=7.0cm]{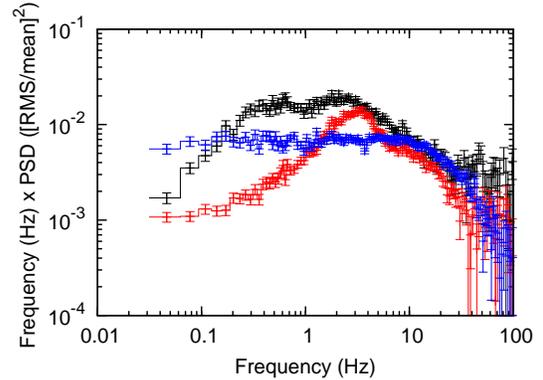} 
\caption{Power spectrum evolution of Cygnus X-1 over an X-ray state change. The source evolves from an intermediate state (black) to a soft state (red) and finally a very soft state (blue) on 2010 June 19, July 3 and July 22 (marked on Fig.~\ref{fig:lightcurve}).} 
\label{fig:XTE_results} 
\end{figure}

The pointed RXTE-PCA observations showed a drop in the RMS variability during the state change (as shown at the top of Fig.~\ref{fig:lightcurve}). The fractional RMS started at $\sim8\%$  on 2010 June 20 before the high-resolution radio monitoring began, and the power spectra (PDS) showed band-limited noise between 0.3--10~Hz (shown in black in Fig.~\ref{fig:XTE_results}), which suggests the source was in an intermediate state \citep{2006ApJ...643.1098S}. The RMS then dropped to $\lesssim4\%$ on 2010 July 4, in between the X-ray and radio flares, and entered a soft state with a somewhat narrower noise component peaking at $\sim3$~Hz (shown in red in Fig.~\ref{fig:XTE_results}). Finally the source entered a very soft state by 2010 July 22 (shown in blue in Fig. 4) with the PDS showing the characteristic broken power-law noise and the RMS remaining $<5\%$ when the VLBI observations detected an unresolved compact jet. Therefore we are confident that Cygnus X-1 entered a soft state during the VLBI observations.

\section{Evidence for a compact jet in a soft state}
\label{compact_jet}
The astrometric residual positions, scattered up and down the jet axis with time in Fig.~\ref{fig:VLBI_results}, are indicative of a compact jet that is slightly smaller than the beam size. The variations in position are likely due to changes in the optical depth ($\tau$) of a partially synchrotron self-absorbed steady jet \citep{1979ApJ...232...34B}. We interpret this as variations in the jet power, electron density or magnetic field causing the opacity to vary and the $\tau=1$ surface to move up and down with time. The larger positional scatter at the longer wavelength also supports the self-absorbed compact jet interpretation, as one expects the apparent size of the outflow to be larger at lower frequencies, since the $\tau=1$ surface is further out. The positions do not evolve linearly with time, and do not correlate with either radio flux density or spectral index.  This suggests a quasi-continuous outflow with variable optical depth rather than a single moving component.

In the hard state, the radio flux is known to be modulated by $\sim15-20\%$ on the 5.6 day orbital period~\citep{1999MNRAS.302L...1P,2006MNRAS.368.1025L}. It has also been suggested that the radio modulation is due to orbital changes in the line-of-sight free-free absorption of the stellar wind from the massive companion star \citep{2002MNRAS.336..699B,2007MNRAS.375..793S}. During the soft state, VLBI observations presented here have shown the jet flux at 4~cm is contained within $\sim0.7$~AU which is approximately an order of magnitude smaller than in the hard state at the same wavelength~\citep{rushton_thesis}. Therefore the soft state quenching of the radio flux could be partly due to an increased absorption by the wind as the emission originates closer to the black hole and companion star. While the sampling in time is too sparse and the signal-to-noise is too low to rule out conclusively that a similar periodicity exists in the soft state, no clear orbital modulation was found during the VLBI results presented in this paper. Therefore any variations are likely to be a combination of optical depth changes in both the jet and stellar wind.

The discovery of a compact jet in the soft state of a BH XRB has important implications for a universal model of accreting black holes. Regardless of whether the X-ray spectrum is dominated by a power-law component or soft thermal emission, a compact jet can remain a fundamental aspect of the accretion, even in a soft state (although the jet may not always be detectable). Across the state transition there is a complex anti-correlation between the soft X-rays and the radio jet, although a deterministic relationship does not always exist; there is a delay between changes in the inflow and outflow, as the soft X-ray flare clearly peaks $\sim3$~days before the radio. However, once the source had entered the soft state for a few days, the radio jet was clearly quenched by a factor of $\sim3-5$ and a few months later reduced by a factor of $>7$. It should however be noted that the soft state of Cygnus X-1 might not be representative of XRBs in general, as the X-ray RMS remains above $\sim4\%$ and may not get to the very low variability seen in the completely radio quenched states discussed by~\cite{2009MNRAS.396.1370F}.

Clear comparisons can be made with the BH candidate XRB GX\,339-4, which also shows a quenching of the radio emission when the source transits into the soft state.  Furthermore, the anti-correlation between the radio and X-ray flux in the soft state with GX\,339-4 is even more pronounced. Whilst the radio flux density became undetectable with ATCA ($<0.1$~mJy) and $>25-40$ times weaker than in the corresponding hard state, the X-ray emission increased by a factor of $\sim10$~\citep{{1999ApJ...519L.165F},2000A&A...359..251C}. Similarly, the BH candidate XRBs H\,1743-322~\citep{2011MNRAS.414..677C} and 4U1957+11~\citep{2011arXiv1106.0723R} have an even larger radio quenching of at least two orders of magnitude in their respective soft states. Therefore if these sources still have a soft-state outflow, they are extremely weak.

\section{Absence of bright discrete ejecta}

The absence of bright, discrete ejections during the transition from the hard to soft state was unexpected. While a (delayed) radio flare was observed shortly after the X-ray state transition, the absence of resolved jet knots suggests either (i) no shocks formed within the compact jet, implying no sudden change in bulk Lorentz factor or (ii) if a knot was produced, it expanded rapidly with a large opening angle and faded very quickly; however, the latter argument is unlikely as the surrounding jet medium is filled with material from the stellar wind that would interact with the knot \cite[N.B. the knot seen by][is possibly an imaging artefact]{2006MNRAS.369..603F}.

These observations therefore demonstrate that not all black hole candidates necessarily produce strong discrete ejecta during a state transition; it was also reported by \cite{2010ATel.2906....1P} that the black hole candidate MAXI J1659-152 did not exhibit any strong discrete ejecta during the early phases of a state change. The reasons for this are unclear. It has been suggested that Cygnus X-1 has a non-spinning black hole \citep{2006ARA&A..44...49R,2009ApJ...697..900M}, which may correspond to an outflow with a much lower velocity than a Kerr black hole \citep[although recently it has been suggested by][that the spin was drastically underestimated]{2011arXiv1106.3690G}. Alternatively, knots could be related to a rapid change in accretion rate ($\dot{m}$), which is not seen in this wind-accreting system with a circular orbit.

Finally the stellar wind itself could be important in suppressing the `jet-line'/formation of knots after a spectral state change. Strong recollimation shocks are thought to occur when the jet interacts with the wind causing disruption even for jet powers of several times $10^{36}$~erg~s$^{-1}$~\citep{2010A&A...512L...4P}. The other BH XRBs discussed in section~\ref{compact_jet} all contain low-mass companions that may only emit an isotropic (disk) wind during the soft state. Therefore, while the quenching of radio emission during the soft state could still be partly due to free-free absorption, high-mass XRBs may always have too dense a surrounding environment to produce knots, unless the jet power is sufficiently high, as may be the case in Cygnus X-3, where discrete ejecta are frequently resolved during radio flares \citep{1988ApJ...331..494M,1995ApJ...447..752S,2001ApJ...553..766M}.

\section{Conclusions}

High-resolution VLBI observations have shown the first evidence of a compact jet-like outflow in a soft state of a BH XRB and no evidence of superluminal motion was detected after a state transition. We have removed all known astrometric signatures and orbital parameters to show that an unresolved compact jet is present, oriented along the same position angle as the larger jet seen in the hard state. Furthermore, we performed a detailed analysis of the X-ray timing properties and hardness ratio over the same time period to confirm the source had entered one of its known soft states; however, we note that during the VLBI observations presented here, Cygnus X-1 may not have entered the canonical soft state (seen in other BH XRBs) that may completely switch off the outflow.

\section{Acknowledgements}
The authors would like to thank the anonymous referee for very useful comments. AR thanks ESO and Marie Curie Actions for a COFUND fellowship. MERLIN is a national facility operated by the University of Manchester and supported by STFC. The European VLBI Network\footnote{\href{http://www.evlbi.org}{www.evlbi.org}} is a joint facility of European, Chinese, South African and other radio astronomy institutes funded by their national research councils. e-VLBI developments in Europe are supported by NEXPReS, an Integrated Infrastructure Initiative (I3), funded under the European Union Seventh Framework Programme (FP7/2007-2013) under grant agreement RI-261525. The NRAO is a facility of the NSF operated under cooperative agreement by Associated Universities, Inc. The WSRT is operated by the ASTRON (Netherlands Institute for Radio Astronomy) with support from the NWO. The AMI is operated by the University of Cambridge and supported by STFC. The RXTE X-ray data was provided by the ASM/\textit{RXTE} teams at MIT and at the \textit{RXTE} SOF and GOF at NASA's GSFC. This work made use of the Swinburne University of Technology software correlator, developed as part of the Australian Major National Research Facilities Programme and operated under licence.

\end{document}